\documentclass[prb,reprint,amsmath,superscriptaddress]{revtex4-1}

\usepackage{graphicx}
\usepackage{times}
\usepackage{bm}

\begin{document}
\title{Dissipation Mechanisms in Thermomechanically Driven Silicon Nitride Nanostrings} 

\author{A.\ Suhel}
\author{B.\ D.\ Hauer}
\author{T.\ S.\ Biswas}
\author{K.\ S.\ D.\ Beach}\email{kbeach@ualberta.ca}
\affiliation{Department of Physics, University of Alberta, Edmonton, Alberta, Canada T6G 2E9}
\author{J.\ P.\ Davis$^{1,}$}\email{jdavis@ualberta.ca}
\affiliation{Canadian Institute for Advanced Research: Nanoelectronics Program, Toronto, Ontario, Canada M5G 1Z8}

\begin{abstract}  
High-stress silicon nitride nanostrings are a promising system for sensing applications because of their ultra-high mechanical quality factors ($Q$s).  By performing thermomechanical calibration across multiple vibrational modes, we are able to assess the roles of the various dissipation mechanisms in these devices.  Specifically, we possess a set of nanostrings in which all measured modes fall upon a single curve of peak displacement versus frequency.  This allows us to rule out bulk bending and intrinsic loss mechanisms as dominant sources of dissipation and to conclude that the most significant contribution to dissipation in high-stress nanostrings occurs at the anchor points.
\end{abstract}

\maketitle

The extremely high values of mechanical $Q$ that have been reported in silicon nitride nanostrings\cite{Ver06,Ver08a,Sch11a} have generated a great deal of excitement in the nanomechanics community.\cite{Unt09}  These devices are ideal for use in mass sensing,\cite{Sch10} temperature sensing,\cite{Lar11} and optomechanics\cite{Eic09,Ane09,Fon10} and 
have proved to be an invaluable platform for research into the quantum properties of nanoscale resonators.\cite{Teu09,Roc10}  Nanostrings possess all the desired properties for these endeavors, including small mass, high frequency, and high $Q$, with correspondingly large displacement amplitudes.  This combination makes them sensitive to external perturbation yet still within the limits of current detection techniques.\cite{Eki05}

In this manuscript, we demonstrate thermomechanically limited detection of up to six mechanical nanostring modes.  Furthermore, we present a set of devices in which all harmonics fall upon a single curve of calibrated peak displacement versus frequency.  As we discuss below, we are able to infer that the mechanical $Q$ is limited by dissipation processes operating in the vicinity of the anchor points---thus suggesting ways to further engineer the mechanical $Q$.

Silicon nitride nanostrings are devices under extremely high tensile stress ($\sigma$ = 0.8 GPa for our devices). 
The accepted understanding is that the tension along their length results in large stored elastic energy and a correspondingly large 
mechanical $Q$.\cite{Sch11a} Experiments have confirmed that $Q$ increases with mechanical tensioning of non-prestressed (low tension, non-stoichiometric silicon nitride) devices,\cite{Ver07,Ver08b} corroborating the tension's central role. 

In addition to causing the high $Q$, the tension compels the devices to behave like strings rather than doubly clamped beams, as one would usually expect for this geometry.  The harmonics are at integer multiples of the fundamental frequency,\cite{Unt10} $\nu_{n}=n\nu_{1}$, where $\nu_{1}$ is the fundamental mode frequency and $n$ indicates the mode number; the mode frequency depends only on the length of the string (not on the width or thickness).  These features are in contrast to the more complicated harmonics of doubly clamped beams,\cite{Cle02} which are realized in low-stress silicon nitride devices of the same geometry.\cite{Sou09}  Furthermore, it has now been shown that very high mechanical $Q$s exist for nanostrings under tension in a variety of other materials, including the polymer SU-8,\cite{Sch08} aluminum,\cite{Lar11,Sul10} and AuPd.\cite{Hua05} This demonstrates that the high $Q$ is a byproduct of the tension and string-like behavior and not a special material property of the silicon nitride itself.  It is likely that most materials that can withstand high tensile stress can be coerced toward higher $Q$ in this way.

\begin{figure}[b]
%%%%%%%%%%%%%%%%%   F I G U R E  1   %%%%%%z%%%%%%%%%%%%
%\centerline{\includegraphics[width=3.1in]{Fig1.png}}
\centerline{\includegraphics[width=3.1in]{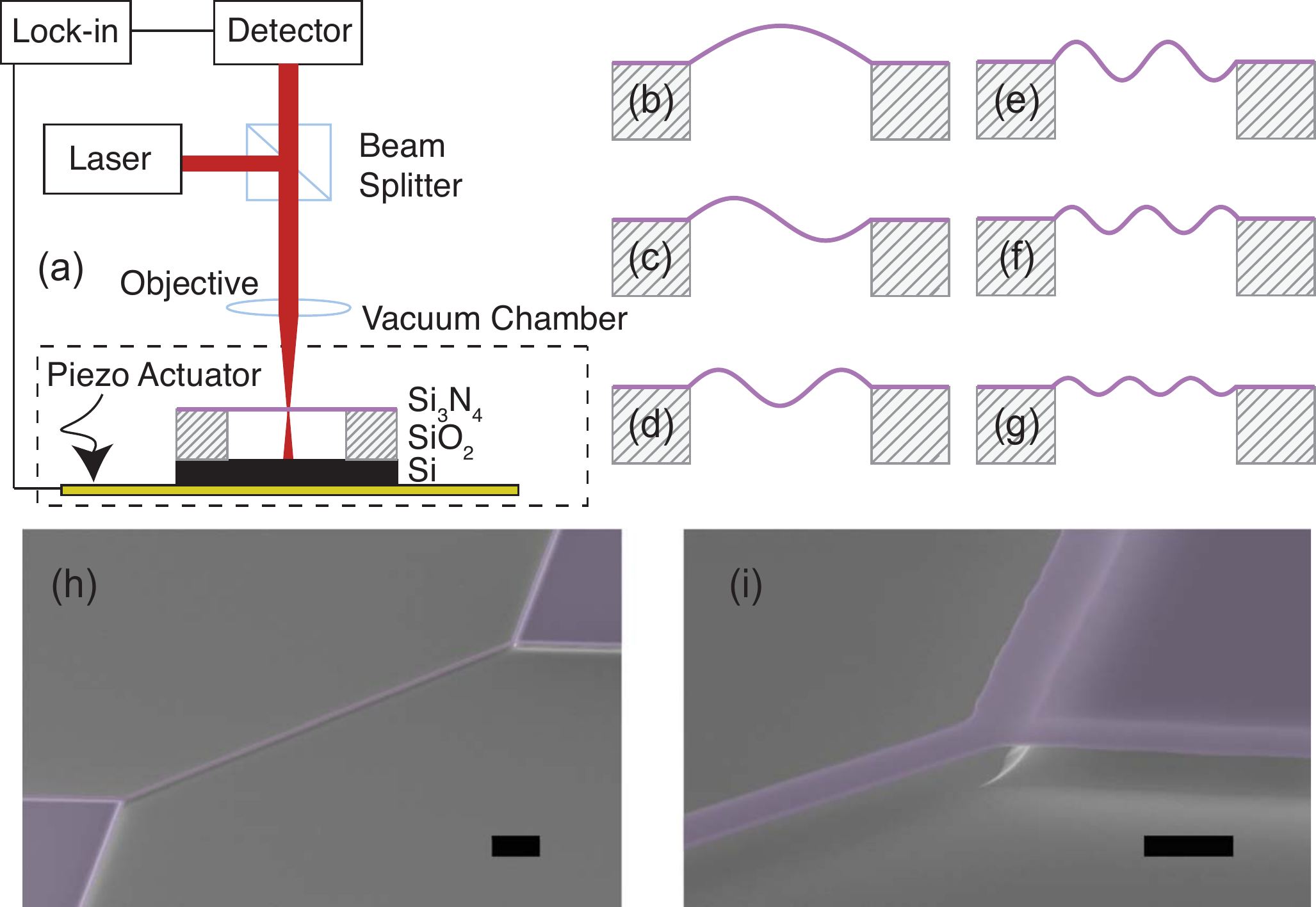}}
%%%%%%%%%%%%%%%%%%%%%%%%%%%%%%%%%%%%%%%%%%%%%
\caption{{\label{fig1}} (a) Schematic of our measurement set-up.  The silicon nitride nanostring is mounted in an optical-access vacuum chamber.  (b--g) Cartoons showing the mode shapes of the first six harmonics of a nanostring.  (h--i) Scanning electron micrographs of the 215~$\mu$m long silicon nitride nanostring, with scale bars of length 20~$\mu$m and 3~$\mu$m respectively. }
\end{figure}

Nanostrings offer significant potential for sensing applications,\cite{Nai09} provided that they can be
accurately calibrated.\cite{Doh07} In this work, the absolute displacement of a nanostring is determined from
measurements of its thermal activation.\cite{Hut93}
 This procedure is often restricted to the fundamental mode, since it is the mode of largest displacement.  Because of the high $Q$ of nanostrings, however, we are able to detect the thermomechanical motion of up to six harmonics for our longest strings. 
 
 A key insight is that such multimode thermomechanical calibration reveals important information about whether the dominant dissipation processes in nanostrings are localized at the anchor points or in the bulk. The calibrated peak displacements of our devices fit an analytical form that follows from very general considerations of the leading order damping allowed by symmetry.  Nanoscale devices may dissipate quite differently from macroscopic oscillators,\cite{Hao03} since these devices have periods of oscillation that are comparable to their thermal transport times and  phonon mean free paths that rival their linear size.  But our analysis is agnostic with respect to the underlying {\it microscopic} mechanism and is consistent with a number of different scenarios for losses at the anchor points (including dissipation by phonon tunneling\cite{Wil08,Wil11} and other models of local bending at the anchor points.\cite{Sch11a,Unt10})

\begin{figure}[t]
%%%%%%%%%%%%%%%%%   F I G U R E  2   %%%%%%%%%%%%%%%%%%
%\centerline{\includegraphics[width=3.4in]{Fig2.png}}
\centerline{\includegraphics[width=3.4in]{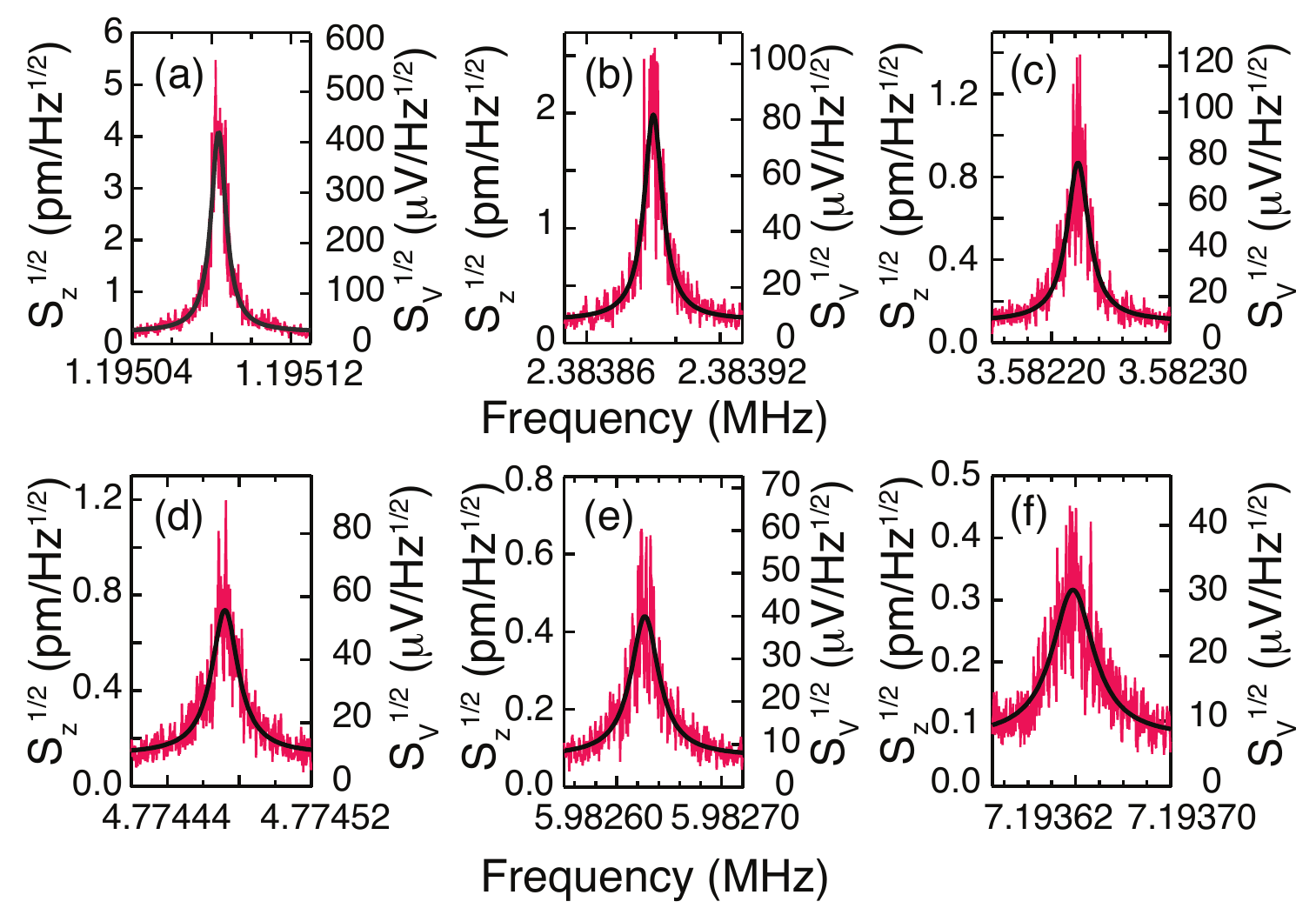}}
%%%%%%%%%%%%%%%%%%%%%%%%%%%%%%%%%%%%%%%%%%%%%
\caption{{\label{fig2}} Calibrated thermomechanical spectra of the first six harmonics of the 215~$\mu$m long nanostring (the device shown in Fig.~\ref{fig1}). These are single sweeps and are post-processed by a sliding average over five points. The data are fit to an analytical expression for the power spectral density function (see Ref.~29.)}
\end{figure}

Our silicon nitride nanostrings are fabricated from stoichiometric silicon nitride deposited onto silicon dioxide on a silicon handle using LPCVD.  This is cost effective as wafers are commercially available for $\approx$\,\$50 for a 100~mm wafer (Rogue Valley Microdevices). We perform standard optical lithography followed by reactive ion etching. The nanostrings are released by a buffered oxide etch of the 2~$\mu$m thick silicon dioxide layer.

The completed devices, shown in Figs.~1(h) and 1(i), are mounted onto a piezoelectric, which can provide mechanical actuation when excited by the lock-in amplifier (Zurich Instruments HF2LI). The experiment is housed in a short-working-distance vacuum chamber at $\approx\! 2\! \times\! 10^{-6}$~torr, which is sufficiently low to eliminate pressure related damping.\cite{Ver08b} The displacement of the nanostring is measured via optical interference with respect to the silicon substrate (HeNe 632.8~nm focused to an $\sim$\! 1~$\mu$m diameter spot), as shown in Fig.~1(a).\cite{Dav10} The 100~$\mu$W of optical power impinging on the string causes no significant heating: we have verified that the resonance frequencies do not shift as a function of optical power,\cite{Sou09} a consequence of the low absorption of silicon nitride in the visible.  Interferometric detection can be performed with or without a voltage applied to the piezo. In the latter case---appropriate for measurement of the thermomechanical actuation---the HF2LI acts as a spectrum analyzer with a 
user-controlled measurement bandwidth.  Because the $Q$s of the devices are so high, a bandwidth of 1 Hz is required. We have identified just how narrow the bandwidth must be by ensuring that the $Q$ measured via thermomechanical actuation is consistent with both the $Q$ of the driven device and the $Q$ extracted from a ringdown measurement versus time.

In a simple model, the string is described by its vertical displacement $z(x,t)$, which is a function of the 
horizontal position $x$ along the string and the time $t$. The string endpoints are fixed at $z(0,t) = z(L,t) = 0$, 
and the string achieves its natural length $L$ when the displacement of the string is everywhere zero. 
In the linear regime---applicable for all measurements reported here---the length of the string deviates from $L$
by an amount $\int_0^L\!dx\,z_x^2$. (Here $z_x = \partial z(x,t)/\partial x$ denotes the partial derivative.) By virtue of the 
strong intrinsic tensile stress $\sigma$ of the silicon nitride, there is a corresponding 
restoring force proportional to the local curvature when the string is stretched. Hence, the
string  obeys a modified wave equation
$\rho z_{tt} = \sigma z_{xx} - D[z(x,t)]$, where the functional $D$ encodes the various dissipative
processes in operation.

It is convenient to decompose the motion of the system into its normal modes
\begin{equation}
z(x,t) = \sum_n a_n(t) \sin\Bigl(\frac{n\pi x}{L}\Bigr),
\end{equation}
which are indexed by a positive integer $n$, the mode number.
The first six mode shapes are sketched in Fig.~1(b)--(g).
Each mode acts as a nearly independent damped harmonic oscillator obeying
\begin{equation} \label{eq:oscillator_ode}
m\frac{d^2 a_{n}}{dt^2}+\gamma_n \frac{\partial a_n}{\partial t}+m\omega^2_na_n = f_n(t).
\end{equation}
Here, $m$ is the geometric mass of the string, $\omega_n$ the angular frequency, and $\gamma_n$ the 
damping coefficient ($\gamma_n = m \omega_n/Q_n$).
In the case of thermomechanical actuation, the forcing term $f_n(t)$ can be understood as a continuous-time
stochastic process that acts as a source of thermal noise.
The sequence of resonance frequencies
\begin{equation} \label{eq:res_freq}
\nu_n = \frac{\omega_{n}}{2\pi}= \frac{n}{2L}\sqrt{\frac{\sigma}{\rho}}
\end{equation}
depends on the device length and on the intrinsic tensile stress and density $\rho$ of the silicon nitride.\cite{Ane09}
Values in the literature for the density vary, and therefore we use our measured frequencies and lengths, along with the tensile stress reported by the manufacturer, $\sigma$ = 0.8 GPa, to determine the density for ourselves. Our estimate of 3000 kg/m$^3$
is consistent with other reported values.\cite{Cla85}

We now briefly discuss the form of the damping coefficient. 
Since the experimental chamber is evacuated to $\approx\! 2\! \times\! 10^{-6}$~torr, the dissipation cannot have a substantial contribution from any viscous term $z_t$ that depends on the absolute motion of the string.\cite{Ver08b} Instead, the leading order contribution from internal material processes must go as ${z_{xx,t}}$,
which describes how fast a volume element in the string is moving with respect to neighboring ones. 
The exception is in the vicinity of the anchor points, where dissipation can depend on the time rate of change of the angle that
the string makes with respect to its connection point ($z_{x,t})$.  By this reasoning, the effective $Q$ for each mode
is well approximated by
\begin{equation} \label{eq:Q_explicit}
Q_n = \frac{m \omega_n}{\gamma_n} \approx \frac{m }{\tilde{\gamma}^\text{visc}\omega_n^{-1}
+ \tilde{\gamma}^\text{anchor} + \tilde{\gamma}^\text{bulk}\omega_n}.
\end{equation}
Here, we have taken
$\gamma_n$ to be a sum of three contributions $\gamma_n^{\text{visc}} = \tilde{\gamma}^\text{visc} \approx 0$,
$\gamma_n^\text{anchor}  = \tilde{\gamma}^\text{anchor} \omega_n \sim \sigma^{-1/2} \omega_n$, and
$\gamma_n^\text{bulk}  = \tilde{\gamma}^\text{bulk} \omega_n^2 \sim \sigma^{-1} \omega_n^2$
that are meant to describe viscous damping, damping at the anchor points, and damping due to bending
of the bulk string. (The tilde-decorated quantities are material constants that 
have had their leading order $\omega_n$ dependence factored out.)
Note that each contribution appears in the denominator of Eq.~\eqref{eq:Q_explicit} with a characteristic mode number (or, equivalently, resonance frequency) dependence, whose exponent $p-1$ tracks the number of $x$ and $t$ derivatives in the relevant dissipation term 
$\partial^{p+1}z/\partial x^p\partial t$. In particular,
$Q$ values that are roughly constant across modes are indicative of dissipation at the anchor points;
dissipation originating deep in the string, on the other hand, should lead to $Q$ that decays inversely with the resonant frequency.
The former behavior is consistent with our understanding that 
the relative contribution $\tilde{\gamma}^\text{bulk} / \tilde{\gamma}^\text{anchor} \sim \sigma^{-1/2}$
is likely to be small in high stress materials.

Thermomechanical calibration allows us to quantitatively compare these loss channels.  Once a thermally driven spectrum is acquired, as in Fig.~2, we calibrate the voltage signal to an absolute displacement via the equipartition theorem.  Fitting the power spectral density function\cite{SupplementaryInfo} to the square of the voltage signal divided by the measurement bandwidth, we extract four parameters: the displacement noise floor, the $Q$, the resonance frequency, and the factor $\alpha$ (having units $m^2/V^2$) that converts between the voltage signal and the displacement. The only uncertainty is the mass of the nanostring, which we compute as $m = \rho Lwt$.  The thickness $t = 250$~nm is the same in all devices, whereas the width $w$ is idiosyncratic.  Accurate knowledge of the density and geometry is important for this calibration.  The main results are the peak displacement of the beam at the resonance frequency (Fig.~3a), and the quality factor $Q_n$ (Fig.~3b), both
of which are extracted from the fit to the calibrated spectrum.  Calibration is repeated for each measurement of a resonance spectrum, since $\alpha$ varies with the location of our laser spot.

The peak displacements and frequencies of the second mode of the 215~$\mu$m device and the first mode of the 102~$\mu$m device are nearly identical [see Fig.~3(a)].  The coincidence of the frequencies (following from the near identity $1/102 \approx 2/215$)
is a consequence of the $n/L$ scaling in Eq.~\eqref{eq:res_freq}. As is characteristic of strings, 
the frequency is proportional to the sound velocity and mode number but otherwise depends only on the string length and not on any of its transverse dimensions.  (The width of the nanostring has been shown to be important when $w>3~\mu\text{m}$,\cite{Sch11a} but when $55~\text{nm} <w<1.5~\mu\text{m}$ the $Q$ is independent of $w$ at low pressures.\cite{Ver08b}) 
There are additional degeneracies---the $3^\text{rd}$ mode of the 215~$\mu$m device and the $1^\text{st}$ mode of the 74~$\mu$m device; the $4^\text{th}$ mode of the 215~$\mu$m device and the $2^\text{nd}$ of the 102~$\mu$m device; the $6^\text{th}$ mode of the 215~$\mu$m device, the $3^\text{rd}$ of the 102~$\mu$m device, and the $2^\text{nd}$ of the 74~$\mu$m device---that together constitute proof of these systems' string-like behavior.

\begin{figure}[t]
%%%%%%%%%%%%%%%%%   F I G U R E  3   %%%%%%%%%%%%%%%%%%
%\centerline{\includegraphics[width=2.5in]{Fig3.png}}
\centerline{\includegraphics[width=2.5in]{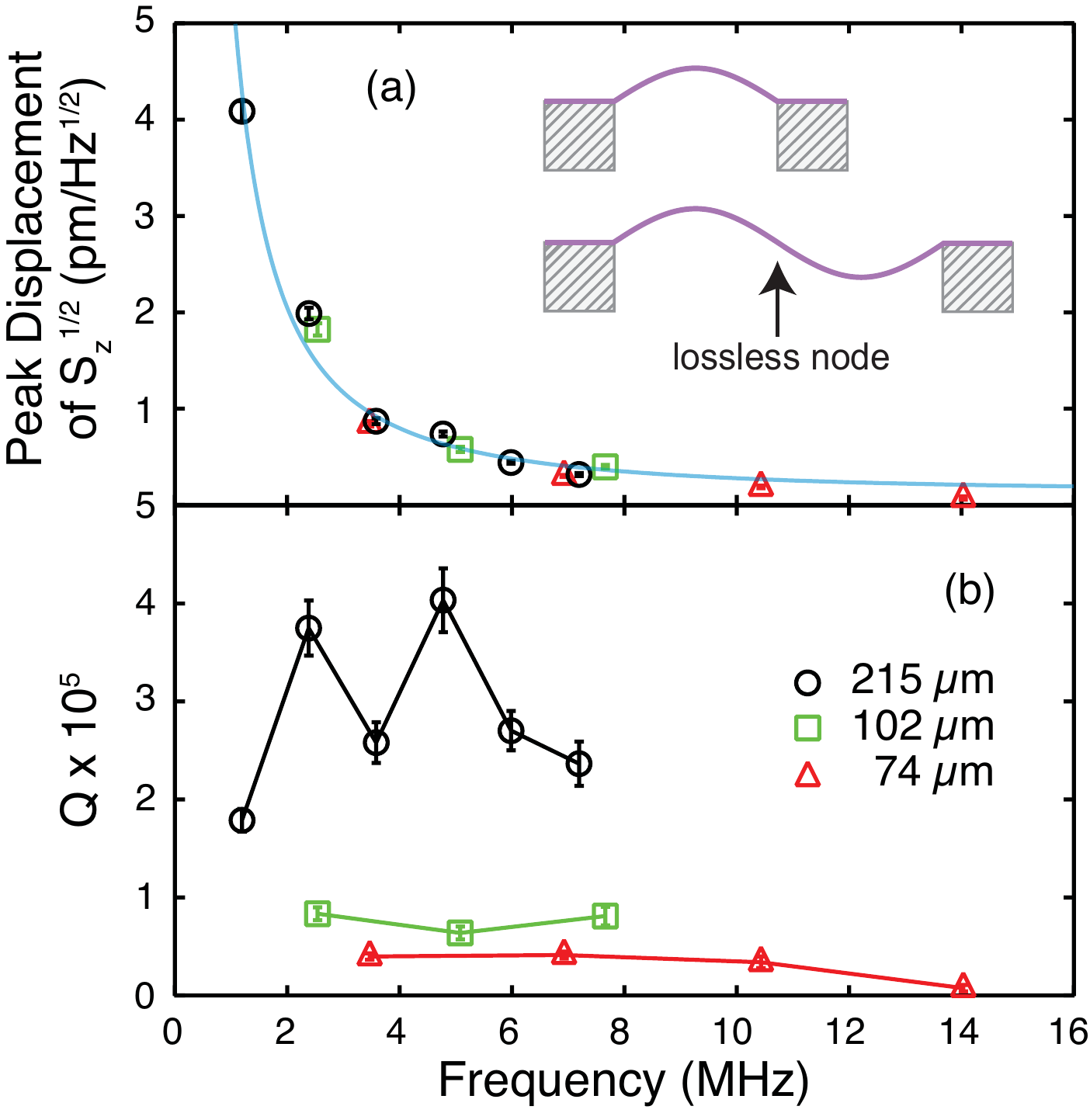}}
%%%%%%%%%%%%%%%%%%%%%%%%%%%%%%%%%%%%%%%%%%%%%
\caption{{\label{fig3}} (a) Peak displacement and (b) $Q$ from thermomechanical calibration of three devices, all 250~nm thick.  Black circles are for a string 215~$\mu$m long and 2.1~$\mu$m wide, green squares for 102~$\mu$m long and 1.1~$\mu$m wide, red triangles for 74~$\mu$m long and 1.2~$\mu$m wide.  The blue curve in (a) is a fit to Eq.~6.}
\end{figure}

On the other hand, it is not obvious that modes with identical frequencies should also have identical peak displacements, 
as seen in Fig.~3(a).  We note this behavior of the peak displacement is not universal. Rather, it is simply fortuitous that we have devices that fall on the same curve, and we have measured other devices that do not fit this trend (for example devices measured at higher pressures).  The fact that these devices were manufactured identically, from the same wafer, and measured under rigorously controlled conditions, aids in the coincidence.  In Fig.~3(b), the harmonics of a longer device are seen to have higher $Q$ (lower dissipation) than the modes of a shorter device with identical frequencies.  Conveniently, there exist in our data set modes from different devices that have nearly identical frequencies and peak displacements but substantially different $Q$s. We must conclude that two such modes differ only in losses at the anchor points. Specifically, since they have identical peak displacements---which is only the case because they are thermomechanically driven---these modes have identical local curvature, and therefore the $Q$ cannot be limited by dissipation due to bulk bending.\cite{Moh05}  We would be unable to make such a claim for a general actuation scenario in which the $Q$ does not depend on amplitude, since both the stored energy and dissipation increase with driven amplitude.  In our thermomechanical scenario, however, the only difference between such modes is the ratio of anchor points (two) to the number of lossless nodes ($n-1$) [see inset to Fig.~3(a)].  The smaller this ratio, the smaller the effective loss at the anchors---dissipation is dominated by the anchor points, be it through local curvature\cite{Sch11a,Unt10} or the tunneling of phonons.\cite{Wil08,Wil11}
 
Moreover, our analytical results tell us that the peak displacement versus frequency curve contains 
quantitative information about the dissipation.\cite{SupplementaryInfo}  In particular, if it is true that loss at the anchor points
dominates, then the peak displacement should decay as $1/\nu_n^{3/2}$. 
A more precise statement of the predicted frequency dependence is as follows:
the spectral density of the squared displacement is
\begin{equation}
S_n(\nu_n^\text{max})=  \frac{Q_n k_BT}{m\pi^3 (1-\frac{1}{4Q_n^2})\nu_n^3}; 
\end{equation}
the corresponding displacement $\Delta z = \sqrt{S_n(\nu_n^\text{max})\Delta\nu}$
over an arbitrary frequency interval $\Delta \nu$ 
goes as
\begin{equation}
\begin{split}
 \sqrt{\frac{Q_n}{\nu_n^3}} 
&\sim \frac{1}{\nu_n^{3/2} \sqrt{\tilde{\gamma}^\text{anchor}}\sqrt{1+(\tilde{\gamma}^\text{bulk}/\tilde{\gamma}^\text{anchor})2\pi\nu_n}}\\
&\sim \frac{1}{\nu_n^{3/2} + \tau\nu_n^{5/2}}.
\end{split}\end{equation}
The fit shown in Fig.~3(a) is of this form, with a value of
 $\tau = \pi \tilde{\gamma}^\text{bulk} / \tilde{\gamma}^\text{anchor}
= (-0.3 \pm 2.5 ) \times 10^{-7}$~s
that is effectively zero within measurement error.
This limit, $\tau^{-1} \gg \nu_n$ for all measured $n$, verifies that the ratio of bulk bending losses to losses at the anchor points is negligible.  This is consistent with the weak variation of $Q$ with respect to mode number that we see in Fig.\ 3b. (When displacements are large, as they are for the 215~nm string, there is an additional even/odd effect related to the parity of the mode shape; this has been shown in high-stress silicon nitride membranes\cite{Wil11}---the 2D equivalent of strings---to result from phonon tunneling through the anchor points.)  Finally, we note that the rate of energy loss is strongly influenced by the design of the junction between the string and the support.\cite{Sch11a,Pho04}  Fitting the calibrated peak displacement versus frequency of the thermomechanical motion, as we have done here, can now be considered a straightforward test of the dominant dissipation and can be applied to other anchor designs.  

Establishing that nanostrings under tensile stress overwhelmingly dissipate from their anchor points opens a path for further increasing $Q$.  Specifically, one could design anchor points at the ends of the beams that either carefully suspend the device\cite{Col11} or reflect phonons back into the string instead of allowing them to leak into the substrate.  This has been achieved in part in Ref.~\onlinecite{Sch11a} by clever fabrication of the nanostrings and sample mounting.  A scheme to eliminate phonon tunneling loss at the anchor points would be to design anchor points with complete phononic bandgaps at the frequencies of the mechanical modes of the nanostrings.  This was recently accomplished for very high frequency modes ($>$ 3 GHz).\cite{Ale11,Cha11}

In conclusion, we have fabricated high $Q$ silicon nitride nanostrings that enable multimode thermomechanical calibration. Measurements of the peak displacement versus frequency for a collection of devices collapse onto a single curve, even though the mechanical $Q$s of these devices do not.  An analysis based on the leading order dissipation allowed by symmetry rules out the possibility that bulk bending and intrinsic loss mechanisms play a leading role.  Instead, we confirm that loss mechanisms at the anchor points dominate. Hence, careful engineering of the anchor points in high stress nanostrings may offer a path to more sensitive devices and new applications.

This work was supported by the University of Alberta, Faculty of Science; the CFI; and NSERC.  We thank Don Mullin, Greg Popowich and Tony Walford for assistance.

\appendix
\section{Supplementary Information}

The string under consideration has a uniform (volumetric) mass density $\rho$
and a uniform cross-sectional area $A$. Even at rest and unstretched,
the string feels a tension $\sigma A$ along its length because of a
stress $\sigma$ that is intrinsic to the material.

Any smooth string profile $z(x)$ between $x=0$ and $x=L$ traces out a total length
\begin{equation}\begin{split} 
\int_0^L \!\sqrt{dx^2 + dz^2} &= \int_0^L \!dx\,\sqrt{1+(dz/dx)^2} \\
&= L + \int_0^L \!dx\,\biggl[ \frac{1}{2}(\partial_x z)^2 - \frac{1}{8}(\partial_x z)^4 + \cdots \biggr].
\end{split}\end{equation}
The string acquires excess potential energy whenever it is stretched beyond its natural length $L$.
For small amplitude oscillations, and in the absence of any explicit dissipation mechanism,
 the energy stored in the string is
\begin{equation} \label{eq:hamiltonian}
H = \int_0^L \! dx\, \biggl[ \frac{1}{2\rho A}\Pi^2 + \frac{A\sigma}{2}(\partial_x z)^2 \biggr]. \end{equation}
Here $\Pi(x)$ is the momentum conjugate to the displacement field.
The relevant boundary conditions are $z(0) = z(L) = 0$. The endpoints
are not effectively clamped, so the gradients $z_x(0)$ and $z_x(L)$ are unrestricted.

(To next leading order Eq.~\eqref{eq:hamiltonian} contains a term proportional to $EA(\partial^2_x z)^2$, where $E$ is the Young's modulus. In a high tensile stress system, the crossover from clamped-beam-like to string-like behave occurs over a small region
$\Delta L \sim \sqrt{E/\sigma} \ll L$. Thus, even when the fourth order term is non-negligible, the string picture is correct in a 
coarse-grained sense.)

Variation of the Hamiltonian by way of the identity
\begin{equation} \frac{\delta}{\delta z(x')} (\partial_x z(x))^2 = - 2\delta(x-x') \partial^2_x z(x) \end{equation}
leads to equations of motion
\begin{equation} \begin{split}
\frac{\delta H}{\delta \Pi(x)} &= \frac{1}{\rho A} \Pi(x) = \dot{z}(x), \\
-\frac{\delta H}{\delta z(x)} &= A \sigma \partial^2_x z(x) = \dot{\Pi}(x) = \rho A \ddot{z}(x).
\end{split} \end{equation}
In other words, the string obeys $\sigma \partial^2_x z =  \rho \ddot{z}$.
Written in terms of the normal modes,
\begin{equation} z(x,t) = \sum_{n=1}^\infty a_n(t) \sin \frac{n \pi x}{L}, \end{equation}
the continuum wave equation reduces to a discrete (but infinite: $n=1,2,\ldots$) set 
of ordinary differential equations
\begin{equation}  \ddot{a}_n + \frac{\sigma}{\rho} \frac{n^2 \pi^2}{L^2} a_n = \ddot{a}_n + \omega_n^2 a_n = 0.\end{equation}
Each normal mode is wholly independent and executes harmonic motion
with angular frequency $\omega_n = (\sigma/\rho)^{1/2} n\pi/L$.

We proceed under the assumption that there are processes that 
allow energy to be shared between all the modes (ergodicity).
Then, if the string is held in thermal equilibrium at temperature $T = 1/k_B \beta$, its properties
are described by a partition function
\begin{equation}\begin{split} Z &= \int \mathcal{D}[z(x)]\mathcal{D}[\Pi(x)]\,e^{-\beta H}\\
&=\int \mathcal{D}[z(x)] e^{-\frac{\beta A\sigma}{2}\int_0^L\!dx\,(\partial_x z)^2 }
\int \mathcal{D}[\Pi(x)] e^{-\frac{\beta}{2\rho A}\int_0^L\!dx\, \Pi^2 },
\end{split}\end{equation}
which takes the form of a Boltzmann-weighted integration over the complete phase space
 of string shapes and momenta.
The $\Pi(x)$ contribution will not enter so long as we only measure
functions of the string displacement. For example,
\begin{equation} \langle z(x)^2 \rangle 
= \frac{\int \mathcal{D}[z(x)] z(x)^2 e^{-\frac{\beta A\sigma}{2}\int_0^L\!dx\,(\partial_x z)^2 }}
{\int \mathcal{D}[z(x)] e^{-\frac{\beta A\sigma}{2}\int_0^L\!dx\,(\partial_x z)^2 }}.
\end{equation}

Once again, we want to decompose the field into its normal modes.
This requires a formal change of variables in the partition function.
We imagine imposing a resolution limit by breaking up the spatial
domain into bins of width $\Delta x = L/N$. At each position
$x_j^* = (j-\tfrac{1}{2})\Delta x$, the displacement field takes a value
coarse-grained $z(x_j^*) \equiv z_j = \sum_{n=1}^N a_n \sin n\pi x_j^*/L$.
The infinitesimals
\begin{equation}dz_j = \sum_{n=1}^N \sin \frac{n\pi(j-\tfrac{1}{2})}{N} da_n \equiv \sum_{n=1}^N J_{j,n} da_n\end{equation}
are connected by a matrix $J$ that itself has no dependence on $\{ z_j \}$
or $\{ a_n \}$. Hence, the Jacobian of the transformation
\begin{equation} \mathcal{D}[z(x)] = \lvert \det J \rvert \prod_{n=1}^N da_n \end{equation}
yields only an inert constant that factors out of any expectation values.

Substituting the normal mode expression into the potential energy term gives
\begin{equation} \begin{split}
\int_0^L\!dx\, \frac{A\sigma}{2} (\partial_x z)^2 
&= \frac{A\sigma}{2} \int_0^L\!dx\,\biggl[ \partial_x \sum_n a_n \sin \frac{n\pi x}{L} \biggr]^2 \\
%&= \frac{A\sigma}{2}\sum_{n,m} a_n a_m \frac{nm\pi^2}{L^2}  \underbrace{\int_0^L\!dx\, \cos \frac{n\pi x}{L}
%\cos \frac{m \pi x}{L}}_{=(L/2)\delta_{n,m}}\\
%&= \frac{A\sigma}{4} \sum_n \frac{n^2\pi^2}{L^2} a_n^2\\
&= \frac{A\rho}{4} \sum_n \omega_n^2 a_n^2
= \frac{m}{4L} \sum_n \omega_n^2 a_n^2.
\end{split}\end{equation}
Here, we've taken advantage of the orthogonality relation
\begin{equation} \int_0^L\!dx\, \cos \frac{n\pi x}{L}
\cos \frac{m \pi x}{L} =\frac{L}{2}\delta_{n,m}. \end{equation}
Gaussian integration of the mode variables gives
\begin{equation} \langle a_n a_m \rangle 
= \frac{\int \bigl[\prod_{n'} a_{n'}\bigr]\,a_na_m\, e^{-\frac{\beta m}{4L}\sum_n \omega_n^2 a_n^2}}
{\int \bigl[\prod_{n'} a_{n'}\bigr]\,e^{-\frac{\beta m}{4L}\sum_n \omega_n^2 a_n^2}}
= \frac{2\delta_{n,m}}{\beta m \omega_n^2},
\end{equation}
which confirms that the thermally averaged potential energy 
\begin{equation} \biggl \langle \int_0^L\!dx\, \frac{A\sigma}{2} (\partial_x z)^2 \biggr\rangle = \frac{Nk_BT}{2}\end{equation}
has $k_B T/2$ per mode.

There's an important subtlety here:
the expectation value
$\langle a_n^2 \rangle = \frac{2k_B T}{m \omega_n^2} $
seems to imply that
$ \frac{1}{2} \omega_n^2 \langle a_n^2 \rangle = k_B T$
is the potential energy in the $n^\text{th}$ oscillator.
It is easier to make the connection with the conventional invocation of the
equipartition theorem (\emph{half} $k_BT$ per degree of freedom)
if we write
\begin{equation} \frac{1}{2}\frac{m}{L} \omega_n^2 \langle a_n^2 \rangle = \frac{1}{2}\tilde{\rho} \omega_n^2 \langle \alpha_n^2 \rangle = \frac{1}{2}k_B T \end{equation}
in terms of the properly normalized mode amplitude $\alpha_n = (2/L)^{1/2} a_n$ (with awkward units
of distance to the half power).
Here, $\tilde{\rho} = m/L$ is the linear mass density.

The average squared displacement of the thermally agitated string is
\begin{equation}\begin{split} 
\langle z(x)^2 \rangle &= \sum_{n,m} \sin \frac{n\pi x}{L} \sin \frac{m \pi x}{L} \langle a_n a_m \rangle \\
&= \sum_n \frac{2 k_B T}{m \omega_n^2} \sin^2 \frac{n\pi x}{L}.
\end{split}\end{equation}
Since the spectrum is harmonic ($\omega_n = n \omega_1$), it is straightforward to evaluate
this expression at the center of the string:
\begin{equation}\begin{split}
\langle z(L/2)^2 \rangle &= \frac{2k_BT}{m \omega_1^2} \sum_{\text{odd $n$}}\frac{1}{n^2}\\
&= \frac{2k_BT}{\rho L} \frac{ \rho L^2}{\sigma \pi^2} \times \frac{\pi^2}{8}\\
&= \frac{k_BTL}{4\sigma}.
\end{split}\end{equation}

We are interested, however, is resolving the motion in the frequency domain.
To do this, we formally identify the thermal and time averaged values
$\langle a_n^2 \rangle_{\text{therm}} = \langle a_n^2 \rangle_{\text{time}}$
and make use of Parseval's theorem to convert to Fourier space,
treating the previous result as 
sum rule on the total square-displacement spectral density:
\begin{equation} \langle a_n^2 \rangle = \frac{1}{t_0}\int_0^{t_0}\!dt\,\lvert a_n(t) \rvert^2 
= \int_0^\infty\!d\omega\, S_n(\omega) =  \frac{2 k_B T}{m \omega_n^2}. \end{equation}

For the undamped oscillator case we have considered so far,
\begin{equation}S_n(\omega) = \frac{2 k_B T}{m \omega_n^2}\delta(\omega-\omega_n),\end{equation}
but we expect the peak to be broadened once dissipation is introduced.
We can solve the relevant damped harmonic oscillator 
\begin{equation} \ddot{a}_n + \frac{\omega_n}{Q_n}\dot{a}_n + \omega_n^2 a_n = \frac{f(t)}{m} \end{equation}
[appearing as Eq.~(2)]
by applying the usual trick of driving it at a single frequency and
complexifying the variable $a_n(t) = \text{Re}\,\tilde{a}_ne^{i\omega t}$.
From
\begin{equation} (-\omega^2\ + i\omega \frac{\omega_n}{Q_n} + \omega_n^2) \tilde{a}_n = \frac{f(\omega)}{m} = C \end{equation}
we get
\begin{equation} a_n = \frac{C \sin(\omega_n t + \phi_n)}{\sqrt{(\omega_n^2-\omega^2)^2 + (\omega \omega_n /Q_n)^2}} \end{equation}
with a phase shift
$\phi_n = \arctan[2\omega\omega_n Q_n^{-1}/(\omega^2-\omega_n^2)]$. The unknown $C$
is constant (independent of $\omega$) because the thermal noise is white. Its value
can be fixed by taking 
the time-averaged
square displacement
\begin{equation} \langle a_n \rangle^2 = \frac{\frac{1}{2}C^2 }{(\omega_n^2-\omega^2)^2 + (\omega \omega_n/Q_n)^2} \end{equation}
and equating the integral
\begin{equation} \int_0^\infty\!d\omega\,\frac{\frac{1}{2}C^2 }{(\omega_n^2-\omega^2)^2 + (\omega \omega_n/Q_n)^2} 
= \frac{1}{2}C^2 \frac{\pi Q_n}{2\omega_n^3}\end{equation}
with the known weight under the curve, $2k_BT/m\omega_n^2$. This
establishes that
\begin{equation} \frac{1}{2}C^2 = \frac{4k_BT}{m\pi Q_n} \end{equation}
and in turn yields
\begin{equation} S_n(\omega)
= \frac{4\omega_n k_BT}{m\pi Q_n[(\omega_n^2-\omega^2)^2 + (\omega \omega_n/Q_n)^2]}. \end{equation}

The corresponding expression in terms of conventional frequency is
\begin{equation} S_n(\nu) = 2\pi S_n(\omega) = \frac{\nu_n k_BT}{m\pi^3 Q_n[(\nu_n^2-\nu^2)^2 + (\nu \nu_n/Q_n)^2]}. \end{equation}
At the peak of the distribution
\begin{equation} \nu^\text{max}_n = \nu_n \sqrt{1-\frac{1}{2Q_n^2}}\end{equation}
the value is
\begin{equation} S_n(\nu_n^\text{max}) =  \frac{Q_n k_BT}{m\pi^3 (1-\frac{1}{4Q_n^2})\nu_n^3}. \end{equation}

In the experimental situation, the relevant quantity is the power spectral density, obtained
by squaring the measured voltage near resonance by the bandwidth of the lock-in amplifier.
It is related to the square-displacement spectral density by
\begin{equation}\begin{split}S_V &= \frac{V^2}{\Delta \nu} = (\text{noise floor}) + \alpha S_z(\nu)\\
&= s_0 + \biggl( \frac{\alpha k_BT}{m\pi^3} \biggr) \frac{\nu_n }{Q_n[(\nu_n^2-\nu^2)^2 + (\nu \nu_n/Q_n)^2]},
\end{split}\end{equation}
where $\alpha$ is a conversion factor having units V$^2/$m$^2$.
It is straightforward to extract $\alpha$ and $s_0$ from a fit to the data.
(There is a unique $\alpha$ associated with each peak.)
Hence, from the measured voltage 
we report a root square-displacement spectral density 
according to
\begin{equation} \sqrt{S_z(\nu)} = \frac{V}{\sqrt{\alpha \Delta \nu}}.\end{equation}

\end{document}